\documentclass[11pt,a4paper,twoside]{article}
\usepackage{graphicx}
\usepackage{color}
\usepackage{hyperref}
\usepackage{setspace}
\usepackage{epstopdf}
\begin{document}
\vspace{20mm}
\baselineskip=0.26in
{\bf \LARGE
\begin{center}
Approximate Analytical Solutions to Relativistic and Nonrelativistic P\"{o}schl-Teller Potential with its Thermodynamic Properties 
\end{center}}
\vspace{2mm}

\begin{center}
{\Large {\bf Sameer M. Ikhdair}}\footnote{\scriptsize E-mail:~ sikhdair@neu.edu.tr;~ sikhdair@gmail.com.} \\
\end{center}
{\small
\begin{center}
{\it Department of Electrical and Electronic Engineering, Near East University, \\
922022 Nicosia, Northern Cyprus, Turkey.} \\
\vspace{2mm}
and\\
\vspace{2mm}
{\it Department of Physics, Faculty of Science, An-Najah National University, \\
New Campus, Nablus, West Bank, Palestine.} 
\end{center}}

\vspace{2mm}
\begin{center}
{\Large {\bf Babatunde J. Falaye}}\footnote{\scriptsize E-mail:~ fbjames11@physicist.net} \\
\end{center}
\begin{center}
{\small
{\it Theoretical Physics Section, Department of Physics, University of Ilorin, \\
 P. M. B. 1515, Ilorin, Nigeria.}}
\vspace{2mm}
\end{center}
\begin{center}
{\Large {\bf Adenike G. Adepoju}}\footnote{\scriptsize E-mail:~adegrace@physicist.net} \\
\small
\vspace{2mm}
{\it Physics Section, Holy Marry College, Ifo, Ogun State, Nigeria.}
\end{center}
\noindent
\begin{abstract}
\noindent
We apply the asymptotic iteration method (AIM) to obtain the solutions of Schr\"{o}dinger equation in the presence of P\"oschl-Teller (PT) potential. We also obtain the solutions of Dirac equation for the same potential under the condition of spin and pseudospin (p-spin) symmetries. We show that in the nonrelativistic limits, the solution of Dirac system converges to that of Schr\"{o}dinger system. Rotational-Vibrational energy eigenvalues of some diatomic molecules are calculated. Some special cases of interest are studied such as $S$-wave case, reflectionless-type potential and symmetric hyperbolic PT potential. Furthermore, we present a high temperature partition function in order to study the behavior of the thermodynamic functions such as the vibrational mean energy $U$, specific heat $C$, free energy $F$ and entropy $S$.															
\end{abstract}


thermodynamic property

\section{Introduction}
The exact solutions of various quantum potential models have attracted much attention from many authors since they contain all necessary information to study quantum models. As well-known, there are several traditional techniques used to solve Schr\"{o}dinger-like differential equation with various quantum potentials \cite{SM1, HA3, ER1, ER2, ER3, ER4}. Few of these mothods include the asymptotic iteration method (AIM) \cite{BJ1, BJ2}, the Nikiforov-Uvarov (NU) method \cite{SM2, HA2, SM3, HA4}. The algebraic techniques are related to the inspection of the Hamiltonian of quantum system as in the supersymmetric quantum mechanics (SUSYQM) \cite{SM4, HA1, HA6} and closely to the factorization method \cite{SM5,SM6}. Other methods are based on the proper and exact quantization rule \cite{SM7,SM8, SM9,SM10} and the SWKB method \cite{SM11}. Except for the previous methods, quasilinearization method (QLM) is dealing with physical potentials numerically \cite{SM12,SM13}.

In the present work, we investigate the Schr\"{o}dinger equation and Dirac equation for the PT potential within the framework of AIM \cite{BJ1, BJ2}. This potential has been investigated by some authors under different wave equations of quantum mechanics which include the Klein-Gordon, the Dirac and the Schr\"{o}dinger equations for the vibrational $\ell=0$ and rotational $\ell\ne0$ states \cite{KG2,KG3,KG4,KG5,KG6,KG7,KG7B,KG8,KG9,DT1,DT2}. The priority purpose for studying this potential is due fact that it has been used to accounted for the physics of many systems which includes the excitons, quantum wires and quantum dots \cite{QD1,QD2,QD3,QD4,QD5,QD6,QD7}. For certain ranges of parameters it behaves like the Kratzer potential. 

In recent years, an asymptotic iteration method for solving second order homogeneous linear differential equations has been proposed \cite{BJ1, BJ2}. This method is a powerful tool in finding the eigensolutions (energy eigenvalues and wave functions) of all solvable quantum potential models \cite{BJ1, BJ2}. This method has been so far applied to solve both the relativistic the non-relativistic quantum mechanical problems \cite{FA1, FA1b, FA1c, FA1d, FA2, FA3}. The purpose of this work is to apply AIM to obtain approximate energy levels and wave functions of the PT potential in the framework Schr\"{o}dinger equation and Dirac equation with the spin and p-spin symmetries by considering an appropriate approximation to the centrifugal (pseudo centrifugal) kinetic energy term and study its thermal properties including vibrational mean energy $U$, specific heat $C$, free energy $F$ and entropy $S$ as given in Ref. \cite{SM14}. Further, the nonrelativistic limit is obtained and some special cases of this potential are investigated.

This paper is organized as follows: In Section 2, we briefly outline the methodology. In Section 3, we present the bound state solutions of the Schr\"{o}dinger  equation with the PT potential. Furthermore, we consider the solution of the PT potential in the framework of the Dirac equation under the spin and p-spin symmetries. The nonrelativistic limit is also obtained. Section 4 presents eigensolutions for some special cases. The thermodynamic properties of the Schr\"{o}dinger equation with PT potential are investigated in section 5. Section 6 is devoted for our numerical results and discussions. Finally we give our conclusion in Section 7.

\section{Method of Analysis}
One of the calculational tools utilized in solving the Schr$\ddot{o}$dinger-like equation including the centrifugal barrier and/or the spin-orbit coupling term is the so called asymptotic iteration method (AIM). For a given potential the idea is to convert the Schr$\ddot{o}$dinger-like equation to the homogenous linear second-order differential equation whose solution is a special function and having the form \cite{BJ1}:
\begin{equation}
y''(x)=\lambda_o(x)y'(x)+s_o(x)y(x),
\label{E1}
\end{equation}
where $\lambda_o(x)$ and $s_o(x)$ have sufficiently many continous derivatives and defined in some interval which are not necessarily bounded. The differential equation (\ref{E1}) has a general solution \cite{BJ1, BJ2}
\begin{equation}
y(x)=\exp\left(-\int^x\alpha(x')dx'\right)\left[C_2+C_1\int^x\exp\left(\int^{x'}\left[\lambda_o(x'')+2\alpha(x'')\right]dx''\right)dx'\right].
\label{E2}
\end{equation}
If $k>0$, for sufficiently large $k$, we obtain the $\alpha(x)$
\begin{equation}
\frac{s_k(x)}{\lambda_k(x)}=\frac{s_{k-1}(x)}{\lambda_{k-1}(x)}=\alpha(x) \ ,\ \ k=1, 2, 3.....
\label{E3}
\end{equation}
where
\begin{eqnarray}
\lambda_k(x)&=&\lambda'_{k-1}(x)+s_{k-1}(x)+\lambda_o(x)\lambda_{k-1}(x),\nonumber\\
s_k(x)&=&s'_{k-1}(x)+s_o(x)\lambda_{k-1}(x) \ ,\ \ k=1, 2, 3.....
\label{E4}
\end{eqnarray}
The energy eigenvalues are obtained from the quantization condition of the method together with equation (\ref{E4}) and can be written as follows:
\begin{equation}
\delta_k(x)=\lambda_k(x)s_{k-1}(x)-\lambda_{k-1}(x)s_{k}(x)=0,\ \  \ k=1, 2, 3....
\label{E5}
  \end{equation}
The energy eigenvalues are then obtained from (\ref{E5}), if the problem is exactly solvable.  If not, for a specific $n$ principal quantum number, we choose a suitable $x_0$ point, determined generally as the maximum value of the asymptotic wave function or the minimum value of the potential and the aproximate energy eigenvalues are obtained from the roots of this equation for sufficiently large values of $k$ with iteration. 
\section{Bound State Solutions}
\subsection{Schr\"{o}dinger equation for PT potential}
The PT potential we shall study is defined as \cite{DT1, PT2, PT3, CS1, SI4}
\begin{equation}
V(r)=\frac{A}{\cosh^2(\alpha r)}+\frac{B}{\sinh^2(\alpha r)},  \  \ \ 0<\alpha r<\frac{\pi}{2}
\label{E6}
\end{equation}
where $A$, $B$ and $\alpha$ are constant coefficients. If we insert this potential into the Schr\"{o}dinger equation, the radial part of the Schr\"{o}dinger equation takes the following form:
\begin{equation}
\frac{d^2R_{n\ell}(r)}{dr^2}+\left[\frac{2\mu}{\hbar^2}E_{n\ell}-\frac{\frac{2\mu}{\hbar^2}A}{\cosh^2(\alpha r)}-\frac{\frac{2\mu}{\hbar^2}B}{\sinh^2(\alpha r)}-\frac{\ell(\ell+1)}{r^2}\right]R_{n\ell}(r)=0,
\label{E7}
\end{equation}
where $n$ and $\ell$ denote the radial and orbital angular momentum quantum numbers, respectively, and $E_{n\ell}$ denote the bound-state energy eigenvalues. It is clear that the above equation cannot be exactly solved for $\ell\neq0$ because of the centrifugal barrier. To obtain the exact bound-state solution, we have to include some approximation to deal with the centrifugal term. It is found that the following \cite{KG7, DT2}
\begin{equation}
\frac{1}{r^2}\approx\alpha^2\left[4d_0+\frac{1}{\sinh^2(\alpha r)}\right],
\label{E8}
\end{equation}
where $d_0=1/12$, is a good approximation to the centrifugal term in short potential range. Now, if we replace the $1/r^2$ with equation (\ref{E8}) and further defining the following notations 
\begin{equation}
A_1=\frac{2\mu A}{\hbar^2},\ \ \ \ \ B_1=\frac{2\mu}{\hbar^2}B+\ell(\ell+1)\alpha^2,\ \ \ \ \ K_1=\frac{2\mu}{\hbar^2}E_{n\ell}-4\ell(\ell+1)\alpha^2d_0,
\label{E9}
\end{equation}
equation (\ref{E7}) can be easily transformed into
\begin{equation}
\frac{d^2R_{n\ell}(r)}{dr^2}+\left[K_1-\frac{A_1}{\cosh^2(\alpha r)}-\frac{B_1}{\sinh^2(\alpha r)}\right]R_{n\ell}(r)=0.
\label{E10}
\end{equation}
In order to solve with the AIM, equation (\ref{E10}) must be transformed into form of equation (\ref{E1}). The reasonable wave function we propose is as follows
\begin{equation}
R_{n\ell}(r)=\cosh^\gamma(\alpha r)\sinh^\beta (\alpha r)F_{n\ell}(r),
\label{E11}
\end{equation}
where $\gamma=\frac{1}{2}\left(1+\sqrt{1-\frac{4A_1}{\alpha^2}}\right)$ and $\beta=\frac{1}{2}\left(1-\sqrt{1+\frac{4B_1}{\alpha^2}}\right)$. On putting this wave function into equation (\ref{E10}), we then arrive at the following second-order homogeneous linear differential equation of the form
\begin{equation}
F_{n\ell}''(r)+2\alpha\left[\frac{\gamma\sinh(\alpha r)}{\cosh(\alpha r)}+\frac{\beta\cosh(\alpha r)}{\sinh(\alpha r)}\right]F_{n\ell}'(r)+K_2F_{n\ell}(r)=0,
\label{E12}
\end{equation}
where we have introduced a new parameter $K_2=K_1+\alpha^2(\beta+\gamma)^2$ for the sake of simplicity. Now, before we apply the  AIM, let us introduce a new variable of the form $z=\sinh(\alpha r)$ in order to avoid the cumbersome calculations. Therefore, equation (\ref{E12}) can be re-written as
\begin{equation}
F_{n\ell}''(z)+\left[\frac{(2\gamma+1)z^2+2\beta(z^2+1)}{z(z^2+1)}\right]F'(z)+\frac{K_2}{\alpha^2(1+z^2)}F(z)=0
\label{E13}
\end{equation}
By making a comparism between equations (\ref{E1}) and (\ref{E13}), we can write $\lambda_0(z)$ and $s_0(z)$ values and be using equation (\ref{E4}), we can calculate $\lambda_k(z)$ and $s_k(z)$ as
\begin{eqnarray}
\lambda_0(z)&=&\frac{(2\gamma+1)z^2+2\beta(z^2+1)}{z(z^2+1)},\nonumber\\ s_0(z)&=&\frac{K_2}{\alpha^2(1+z^2)},\nonumber\\
\lambda_1(z)&=&2z^2\frac{2\gamma+1}{(z^2+1)^2}+2\frac{\beta}{z^2}-\frac{2\gamma+1}{z^2+1}-\frac{K_2}{\alpha^2(z^2+1)}+\left(z\frac{2\gamma+1}{z^2+1}+2\frac{\beta}{z}\right)^2,\nonumber\\
s_1(z)&=&K_2\frac{(3+2\beta+2\gamma)z^2+2\beta}{\alpha^2z(1+z^2)},\nonumber\\
\ldots etc.
\label{E14}
\end{eqnarray}
By using the quantization condition given by equation (\ref{E5}), we can establish the following relations
\begin{eqnarray}
\delta_1&=&0\ \ \ \ \ \Rightarrow\ \ \ \ \ K_2=0,\nonumber\\
\delta_2&=&0\ \ \ \ \ \Rightarrow\ \ \ \ \ K_2=-4\alpha^2(\gamma+\beta+1),\nonumber\\
\delta_3&=&0\ \ \ \ \ \Rightarrow\ \ \ \ \ K_2=-4\alpha^2(\gamma+\beta+2),\nonumber\\
\delta_4&=&0\ \ \ \ \ \Rightarrow\ \ \ \ \ K_2=-4\alpha^2(\gamma+\beta+3),\nonumber\\
\ldots etc.
\label{E15}
\end{eqnarray}
We generalize the above expression by induction and substituting for $K_2$, the eigenvalues becomes
\begin{equation}
K_1+\alpha^2(\gamma+\beta+2n)^2=0,\ \ \ \ \ \ \ \ \ \ \ \ n=0, 1, 2,\ldots.
\label{E16}
\end{equation}
By using the notations in equation (\ref{E9}), we obtain a more explicit expression for the energy eigenvalue equation as:
\begin{equation}
E_{n\ell}=\frac{2\alpha^2\hbar^2}{\mu}\left[\ell(\ell+1)d_0-\left(n+\frac{1}{2}+\frac{1}{4}\sqrt{1-\frac{8\mu A}{\alpha^2\hbar^2}}-\frac{1}{4}\sqrt{(2\ell+1)^2+\frac{8\mu B}{\alpha^2\hbar^2}}\right)^2\right].
\label{E17}
\end{equation}
Now, let us study the eigenfunction of the system. To perform this task, the differential equation we wish to solve should be transformed to the form \cite{BJ2}: 
\begin{equation}
y^{\prime \prime }(x)=2\left( \frac{\Lambda x^{N+1}}{1-bx^{N+2}}-\frac{m+1}{x}\right) y^{\prime }(x)-\frac{Wx^{N}}{1-bx^{N+2}}, 
\label{E18}
\end{equation}
where $\Lambda$, $b$ and $m$ are constants. The general solution of equation (\ref{E18}) can be found as {\cite{BJ2}} 
\begin{equation}
y_{n}(x)=(-1)^{n}C_{2}(N+2)^{n}(\sigma )_{_{n}}{_{2}F_{1}(-n,t+n;\sigma;bx^{N+2})},  
\label{E19}
\end{equation}
where the following parameters have been used 
\begin{equation}
(\sigma )_{_{n}}=\frac{\Gamma {(\sigma +n)}}{\Gamma {(\sigma )}}\ \ ,\ \
\sigma =\frac{2m+N+3}{N+2}\ \ and\ \ \ t=\frac{(2m+1)b+2\Lambda }{(N+2)b}.
\label{E20}
\end{equation}
By comparing equations (\ref{E19}) with (\ref{E13}), we can easily determine the parameters $\Lambda$, $b$, $N$, $m$ and $\sigma$. Consequently, the radial eigenfunction for the Schr\"{o}dinger equation with the PT potential can be found as
\begin{equation}
R_{n\ell}(r)=2^n(-1)^nC_2\cosh^\gamma(\alpha r)\sinh^\beta(\alpha r)\frac{(\beta+n)!}{\beta!}\ {_2F_1}\left(-n,\beta+\gamma+n;\beta+1;-\sinh(\alpha r)\right),
\label{E21}
\end{equation}
where $C_2$ is the normalization factor.
\subsection{Dirac equation for the PT potential}
Using Dirac wave equation and Dirac spinor wave functions, the two-coupled
second-order ordinary differential equations for the upper and lower components of the
Dirac wave function can be obtained as {\cite{DT1,DT2,SI1,SI2,SI3,SI4}}
\begin{equation}
\left(\frac{d}{dr}+\frac{\kappa}{r}\right)F_{n\kappa}(r)=\left[M+E_{{n\kappa}}-\Delta(r)\right]G_{n\kappa}(r),
\label{E22}
\end{equation}
\begin{equation}
\left(\frac{d}{dr}-\frac{\kappa}{r}\right)G_{n\kappa}(r)=\left[M-E_{{n\kappa}}+\Sigma(r)\right]F_{n\kappa}(r).
\label{E23}
\end{equation}
where $\Delta(r)=V(r)-S(r)$ and $\Sigma(r)=V(r)+S(r)$. On solving equation (\ref{E22}), we obtain the following Schr\"{o}diger-like differential equation with coupling to the $r^{-2}$ singular term and satisfying $F_{n\kappa}(r)$:
\begin{equation}
\left[\frac{d^2}{dr^2}-\frac{\kappa(\kappa+1)}{r^2}+\frac{\frac{d\Delta(r)}{dr}\left(\frac{d}{dr}+\frac{\kappa}{r}\right)}{M+E_{n\kappa}-\Delta(r)}
-\left(M+E_{n\kappa}-\Delta(r)\right)\left(M-E_{n\kappa}+\Sigma(r)\right)\right]F_{n\kappa}(r)=0,
\label{E24}
\end{equation}
where $E_{n\kappa}\neq-M$ when $\Delta(r)=0$ and $\kappa(\kappa+1)=\ell(\ell+1)$. Since $E_{n\kappa}=+M$ is an element of the positive energy spectrum of the Dirac Hamiltonian, this relation with upper spinor component is not valid for the negative energy spectum solution. Furthermore, a similar equation satisfying $G_{n\kappa}(r)$ can be obtained as:
\begin{equation}
\left[\frac{d^2}{dr^2}-\frac{\kappa(\kappa-1)}{r^2}+\frac{\frac{d\Sigma(r)}{dr}\left(\frac{d}{dr}-\frac{\kappa}{r}\right)}{M-E_{n\kappa}+\Sigma(r)}
-\left(M+E_{n\kappa}-\Delta(r)\right)\left(M-E_{n\kappa}+\Sigma(r)\right)\right]G_{n\kappa}(r)=0,
\label{E25}
\end{equation}
where $E_{n\kappa}\neq+M$ when $\Sigma(r)=0$ and $\kappa(\kappa-1)=\tilde{\ell}(\tilde{\ell}+1)$. Since $E_{n\kappa}=-M$ is an element of the negative energy spectrum of the Dirac Hamiltonian, this relation with the lower spinor component is not valid for the positive energy spectrum solution. In the next subsections, we shall study the PT potential within the framework the Dirac theory in the presence of spin and p-spin symmetries.
\subsubsection{Bound states in pseudospin symmetry limit}
Under the condition of the p-spin symmetry, i.e. ,$\frac{d\Sigma(r)}{dr}=0$, equation (\ref{E25}) turns to
\begin{equation}
\left[\frac{d^2}{dr^2}-\frac{\kappa(\kappa-1)}{r^2}-\left(M+E_{n\kappa}-\Delta(r)\right)\left(M-E_{n\kappa}+C_{ps}\right)\right]G_{n\kappa}(r)=0.
\label{E26}
\end{equation}
Now taking $\Delta(r)$ as the PT potential, using approximation expression (\ref{E8}), defining the following parameters
\begin{eqnarray}
\tilde{A_3}&=&\left(E_{n\kappa}-M-C_{ps}\right)\frac{A}{\alpha^2}, \ \tilde{B_3}=\left(E_{n\kappa}-M-C_{ps}\right)\frac{B}{\alpha^2}+\kappa(\kappa-1),\nonumber\\ \tilde{K_3}&=&4\kappa(\kappa-1)d_0+\frac{1}{\alpha^2}\left(M-E_{n\kappa}+C_{ps}\right)\left(M+E_{n\kappa}\right),
\label{E27}
\end{eqnarray}
and introducing a new transformation of the form $z=\cosh^2(\alpha r)$, equation (\ref{E26}) can be transformed easily to
\begin{equation}
G_{n\kappa}''(z)+\left[\frac{1-2z}{2z(1-z)}\right]G_{n\kappa}'(z)+\frac{1}{4z(1-z)}\left[\frac{\tilde{A_3}}{z}-\frac{\tilde{B_3}}{1-z}+\tilde{K_3}\right]G_{n\kappa}(z)=0.
\label{E28}
\end{equation}
In order to solve this equation with AIM, the wave function sitisfies the boundary conditions is being proposed as
\begin{equation}
G_{n\kappa}(z)=(1-z)^{\tilde{\gamma_2}} z^{\tilde{\beta_2}}g_{n\kappa}(z),
\label{E29}
\end{equation}
where $\tilde{\gamma_2}=\frac{1}{4}\left(1-\sqrt{1+4\tilde{B_3}}\right)$ and $\tilde{\beta_2}=\frac{1}{4}\left(1-\sqrt{1-4\tilde{A_3}}\right)$. Inserting this wave function into equation (\ref{E28}), leads us to write
\begin{equation}
g_{n\kappa}''(z)+2\left(\frac{\tilde{\beta_2}+\frac{1}{4}}{z(1-z)}-\frac{\tilde{\beta_2}+\tilde{\gamma_2}+\frac{1}{2}}{1-z}\right)g_{n\kappa}'(z)+\left(\frac{\tilde{K_3}-4(\tilde{\beta_2}+\gamma_1)^2}{4z(1-z)}\right)g_{n\kappa}(z)=0.
\label{E30}
\end{equation}
Following same procedures as in the previous section, we can use equation (\ref{E30}) to calculate the $\lambda_k(z)$ and $s_k(z)$ then combine the result with the quantization condition given by equation (\ref{E5}). This yields
\begin{eqnarray}
s_0\lambda_1-s_1\lambda_0&=&0\ \ \ \ \ \ \ \ \ \ \Rightarrow\ \ \ \ \ \ \ \ \ \ \tilde{A_3}=4(\tilde{\gamma_2}+\tilde{\beta_2})^2\nonumber\\
s_1\lambda_2-s_2\lambda_1&=&0\ \ \ \ \ \ \ \ \ \ \Rightarrow\ \ \ \ \ \ \ \ \ \ \tilde{A_3}=4(\tilde{\gamma_2}+\tilde{\beta_2}+1)^2\nonumber\\
s_2\lambda_3-s_3\lambda_2&=&0\ \ \ \ \ \ \ \ \ \ \Rightarrow\ \ \ \ \ \ \ \ \ \ \tilde{A_3}=4(\tilde{\gamma_2}+\tilde{\beta_2}+2)^2\nonumber\\
\ldots etc.
\label{E31}
\end{eqnarray}
By generalizing the above expression and using the notations in equation (\ref{E27}), the relativistic energy spectrum becomes
\begin{eqnarray}
&&4\alpha^2\left[d_0\kappa(\kappa-1)-\left[n+\frac{1}{2}+\frac{1}{4}\sqrt{1+\frac{4A}{\alpha^2}\left(M-E+C_{ps}\right)}-\frac{1}{4}\sqrt{(2\kappa-1)^2-\frac{4B}{\alpha^2}\left(M-E_{n\kappa}+C_{ps}\right)}\right]^2\right]\nonumber\\
&&+\left(M+E_{n\kappa}\right)\left(M-E_{n\kappa}+C_{ps}\right)=0,
\label{E32}
\end{eqnarray}
which is identical to Ref. \cite{DT2}  if $B \rightarrow \alpha ^{2} B(B-\alpha)/(2M)$ and $A \rightarrow -\alpha ^{2} A(A+\alpha)/(2M)$. Let us now turn to the calculations of the corresponding wave functions for this system. Thus using the procedure described by equation (\ref{E18})-(\ref{E20}), the wave functions can be easily found as
\begin{equation}
G_{n\kappa}(r)=\tilde{N}_{n\kappa}\frac{\Gamma\left(2\tilde{\beta_2}+n+\frac{1}{2}\right)}{\Gamma\left(2\tilde{\beta_2}+\frac{1}{2}\right)}\cosh^{2\tilde{\beta_2}}(\alpha r)\sinh^{2\tilde{\gamma_2}}(\alpha r) \ _2F_1\left(-n,2(\tilde{\beta_2}+\tilde{\gamma_2})+n;2\tilde{\beta_2}+\frac{1}{2};\sinh^2(\alpha r)\right),
\label{E33}
\end{equation}
where $\tilde{N}_{n\kappa}$ is the normalization constant.
\subsubsection{Bound states in spin symmetry liimit}
Under the condition of the spin symmetry, i.e. ,$\frac{d\Delta(r)}{dr}=0$, equation (\ref{E24}) reduces to
\begin{equation}
\left[\frac{d^2}{dr^2}-\frac{\kappa(\kappa+1)}{r^2}-\left(M-E_{n\kappa}+\Sigma(r)\right)\left(M+E_{n\kappa}-C_{s}\right)\right]F_{n\kappa}(r)=0.
\label{E34}
\end{equation}
Now taking the sum potential as the PT potential along with the approximation expression given by equation (\ref{E8}) and then introducing a new parameter of the form $z(r)=\sinh(\alpha r)$, the equation (\ref{E26}) can be easily decomposed into a Schr\"{o}dinger-like equation in the spherical coordinates for the upper-spinor component $F_{n\kappa}(r)$,
\begin{equation}
F_{n\kappa}''(z)+\left[\frac{1-2z}{2z(1-z)}\right]F_{n\kappa}'(z)+\frac{1}{4z(1-z)}\left[\frac{A_3}{z}-\frac{B_3}{1-z}+K_3\right]F_{n\kappa}(z)=0,
\label{E35}
\end{equation}
where
\begin{eqnarray}
A_3&=&\left(E_{n\kappa}+M-C_{s}\right)\frac{A}{\alpha^2}, \ B_3=\left(E_{n\kappa}+M-C_{s}\right)\frac{B}{\alpha^2}+\kappa(\kappa+1),\nonumber\\ K_3&=&4\kappa(\kappa+1)+\frac{1}{\alpha^2}\left(M+E_{n\kappa}-C_{s}\right)\left(M-E_{n\kappa}\right),
\label{E36}
\end{eqnarray}
In order to avoid repetition of algebral, a first inspection for the relationship between the present set of parameters ($A_3$, $B_3$, $K_3$) and the previous set ($\tilde{A_3}$, $\tilde{B_3}$, $\tilde{K_3}$) enable us to know that the positive solution for the above equation (\ref{E35}), can be easily obtain by using the parameter parameter map \cite{BJ3, BJ4}
\begin{eqnarray}
G_{n\kappa}\leftrightarrow F_{n\kappa},\ \ \ V(r)\rightarrow-V(r),\ \ \ E_{n\kappa}\rightarrow-E_{n\kappa},\ \ \ \kappa \rightarrow \kappa+1 \ \ \ and \ \ \ C_{ps}\rightarrow-C_{s}.
\label{E37}
\end{eqnarray}
Using the above transformations and following the previous results, we obtain the relativistic energy spectrum as
\begin{eqnarray}
&&4\alpha^2\left[d_0\kappa(\kappa+1)-\left[n+\frac{1}{2}+\frac{1}{4}\left(\sqrt{1-4A\frac{M+E-C_{s}}{\alpha^2}}-\sqrt{(2\kappa+1)^2+\frac{4B}{\alpha^2}\left(M+E_{n\kappa}-C_{s}\right)}\right)\right]^2\right]\nonumber\\
&&+\left(M-E_{n\kappa}\right)\left(M+E_{n\kappa}-C_{s}\right)=0,
\label{E38}
\end{eqnarray}
which is identical to Ref. \cite{DT2}  if $B \rightarrow \alpha ^{2} B(B-\alpha)/(2M)$ and $A \rightarrow -\alpha ^{2} A(A+\alpha)/(2M)$, and the corresponding wave functions as
\begin{equation}
F_{n\kappa}(r)={N_{n\kappa}}\frac{\Gamma\left(2{\beta_2}+n+\frac{1}{2}\right)}{\Gamma\left(2{\beta_2}+\frac{1}{2}\right)}\cosh^{2{\beta_2}}(\alpha r)\sinh^{2{\gamma_2}}(\alpha r) \ _2F_1\left(-n,2({\beta_2}+{\gamma_2})+n;2{\beta_2}+\frac{1}{2};\sinh^2(\alpha r)\right),
\label{E39}
\end{equation}
where $N_{n\kappa}$ is the normalization constant.
\section{Some special cases of the PT potential}
In this section, we shall study four special cases of the energy eigenvalues (\ref{E32}) and (\ref{E38}) for the p-spin and spin symmetry, respectively.
\subsection{S-wave case}
By considering the case $\tilde{\ell}=0$ $(\kappa=1)$ and $\ell=0$ $(\kappa=-1)$, the relativistic energy spectra for the exact p-spin ($C_{ps}=0$) and exact spin  symmetry ($C_{s}=0$) become
\begin{eqnarray}
M^2-E^2_{n,+1}=4\alpha^2\left[n+\frac{1}{2}+\frac{1}{4}\sqrt{1+\frac{4A}{\alpha^2}\left(M-E_{n,+1}\right)}-\frac{1}{4}\sqrt{1-\frac{4B}{\alpha^2}\left(M-E_{n,+1}\right)}\right]^2,
\label{E40}
\end{eqnarray}
and
\begin{eqnarray}
M^2-E^2_{n,-1}=4\alpha^2\left[n+\frac{1}{2}+\frac{1}{4}\sqrt{1-\frac{4A}{\alpha^2}\left(M+E_{n,-1}\right)}-\frac{1}{4}\sqrt{1+\frac{4B}{\alpha^2}\left(M+E_{n,-1}\right)}\right]^2,
\label{E41}
\end{eqnarray}
respectively. Equation (\ref{E41}) is the same as the energy equation obtained by equation (22) of Ref. \cite{PT2} and (35) of Ref. \cite{PT3} by taking $A\rightarrow -2A(A+\alpha)$, $B\rightarrow 2B(B-\alpha)$.
\subsection{Reflectionless-type potential}
Choosing $B=0$, $A=-\eta(\eta+1)/2$ in potential (\ref{E1}) then it becomes the reflectionless-type potential \cite{BJ5, BJ6}
\begin{equation}
V(r)=-\frac{\eta(\eta+1)}{2\cosh^2(\alpha r)},
\label{E42}
\end{equation} 
where $\eta$ is an integer. The eigenvalue equation for exact p-spin and spin symmetry become
\begin{eqnarray}
M^2-E^2_{n,+1}=4\alpha^2\left[n+\frac{1}{4}+\frac{1}{4}\sqrt{1-\frac{2\eta(\eta+1)}{\alpha^2}\left(M-E_{n,+1}\right)}\right]^2,
\label{E43}
\end{eqnarray}
and
\begin{eqnarray}
M^2-E^2_{n,-1}=4\alpha^2\left[n+\frac{1}{4}+\frac{1}{4}\sqrt{1+\frac{2\eta(\eta+1)}{\alpha^2}\left(M+E_{n,-1}\right)}\right]^2,
\label{E44}
\end{eqnarray}
respectively.
\subsection{Symmetric hyperbolic modified PT potential}
On puting $\alpha=1$, $B=0$, and $A=\frac{1}{4}-\eta^2$ in the potential equation (\ref{E1}), we have \cite{BJ5, BJ6}
\begin{equation}
V(r)=\frac{\frac{1}{4}-\eta^2}{\cosh^2(r)},
\label{E45}
\end{equation}
with the relativistic energy equation 
\begin{eqnarray}
M^2-E^2_{n,+1}=4\left[n+\frac{1}{4}+\frac{1}{4}\sqrt{1+(1-4\eta^2)\left(M-E_{n,+1}\right)}\right]^2,
\label{E46}
\end{eqnarray}
and
\begin{equation}
M^2-E^2_{n,-1}=4\left[n+\frac{1}{4}+\frac{1}{4}\sqrt{1-(1-4\eta^2)\left(M+E_{n,-1}\right)}\right]^2,
\label{E47}
\end{equation}
for p-spin and spin symmetries, respectively.
\subsection{The non-relativistic limit}
As it can be seen the nonrelativistic Schr\"{o}dinger equation is bosonic in nature, i.e., spin does not involve in it. On the other hand, relativistic Dirac equation is for a spin$-1/2$ particles. It implicitly suggests that there may be a certain relationship between the solutions of the two fundamental equations \cite{BJ7}. de Souza Dutra et al \cite{BJ7} also noted that there is possibility of obtaining approximate nonrelativistic (NR) solutions from relativistic (R) ones. Very recently, H. Sun \cite{BJ7} proposed a little bit crude but meaningful approach for deriving the bound state solutions of NR Schrödinger equation (SE) from the bound state of R equations. The essence of the approach was that, in NR limit, the SE may be derived from the R one when the energies of the two potential $S(r)$ and $V(r)$ are small compared to the rest mass $mc^2$, then the NR energy approximated as $E^{NR}\rightarrow E-mc^2$ and NR wave function is the $\psi^{NR} (r)\rightarrow\psi(r)$. That is, its NR energies, $E^{NR}$ can be determined by taking the NR limit values of the R eigenenergies $E$. 
Therefore, taking $C_s=0$, and using the following transformations $M+E_{n\kappa}\rightarrow 2\mu$ and $M-E_{n\kappa}\rightarrow-E_{n\ell}$ together with $\kappa \rightarrow \ell$ \cite{BJ7}, the relativistic energy equation (\ref{E38}) reduces to
\begin{equation}
E_{n\ell}=\frac{2\alpha^2}{\mu}\left[\ell(\ell+1)d_0-\left(n+\frac{1}{2}+\frac{1}{4}\sqrt{1-\frac{8\mu A}{\alpha^2}}-\frac{1}{4}\sqrt{(2\ell+1)^2+\frac{8\mu B}{\alpha^2}}\right)^2\right],
\label{E48}
\end{equation}
where $\hbar=1$. Let us remark that the non-relativistic solution is identical with the one we obatained for Schr\"{o}dinger case in equation (\ref{E17}). Similarly, we find the non-relativistic of equations (\ref{E44}) and (\ref{E47}) as
\begin{equation}
E_{n}=-\frac{2\alpha^2}{\mu}\left[n+\frac{1}{4}+\frac{1}{4}\sqrt{1+4\mu\eta\frac{(\eta+1)}{\alpha^2}}\right]^2,
\label{E49}
\end{equation}
and
\begin{equation}
E_{n}=-\frac{2}{\mu}\left[n+\frac{1}{4}+\frac{1}{4}\sqrt{1-2\mu\left(1-4\eta^2\right)}\right]^2,
\label{E50}
\end{equation}
for reflectionless-type and symmetric hyperbolic modified PT potential, respectively. It should be noted that equations (\ref{E49}) and (\ref{E50}) can be obtained as special cases of equation (\ref{E17}).
\section{Thermodynamic properties of the Schr\"odinger-PT problem}
In this section, we study the thermodynamic properties of the PT potential model. The eigenvalues $E_{n\ell}$ of this system that we obtained by equation (\ref{E17}) can be re-written as
\begin{equation}
E_{n\ell}=\frac{2\alpha^2\hbar^2}{\mu}\left[\ell(\ell+1)d_0-\left[n-\zeta\right]^2\right], \ \ \ \ n=0, 1, 2, ......<n_{max}=[\zeta],
\label{E51}
\end{equation}
where we have introduced $\zeta=\frac{1}{4}\sqrt{1+\frac{8\mu B}{\alpha^2\hbar^2}}-\frac{1}{4}\sqrt{1-\frac{8\mu A}{\alpha^2\hbar^2}}-\frac{1}{2}+\sqrt{\ell(\ell+1)d_0}$ for mathematical simplicity and $[\zeta]$ means the largest integer inferior to $\zeta$. Secondly, we obtain the vibrational partition function calculated by
\begin{equation}
Z(\beta,\zeta)=\sum^\zeta_{n=0}e^{-\beta E_{n\ell}}\ , \ \ \ \beta=\frac{1}{kT},
\label{E52}
\end{equation}
where $k$ is the Boltzman constant. Now, the substitution of equation (\ref{E51}) into equation (\ref{E52}) yields:
\begin{equation}
Z(\beta,\zeta)=e^{-\frac{2\alpha^2\hbar^2}{\mu}\ell(\ell+1)\beta d_0}\sum^\zeta_{n=0}e^{\left(\frac{n-\zeta}{\gamma}\right)^2}, \ \ \ \gamma=\frac{\tau}{\sqrt \beta}, \  \  \tau=\frac{1}{\alpha\hbar} \sqrt \frac{\mu}{2}, 
\label{E53a}
\end{equation}
Since $e^{-\frac{2\alpha^2\hbar^2}{\mu}\ell(\ell+1)\beta d_0}\approx 1$, the partition function $Z$ can therefore be witten as
\begin{equation}
Z(\beta,\zeta)=\sum^\zeta_{n=0}e^{\left(\frac{n-\zeta}{\gamma}\right)^2}.
\label{E53b}
\end{equation}
In the classical limit, at high temperature $T$ for large $\zeta$ and small $\beta$, the sum can be replaced by the following integral
\begin{eqnarray}
Z(\beta,\zeta)&=&\gamma \int^{\frac{\zeta}{\gamma}}_{0}e^{y^2}dy=\frac{ \sqrt{\pi} \tau \mbox{Erfi} \left(\frac{\zeta}{\tau} \sqrt{\beta} \right)}{2 \sqrt{\beta}},  \ \ \ \ \ \ \ \ y=\frac{\left(n-\zeta\right)}{\gamma}, 
\label{E54}
\end{eqnarray}
Having determined the vibrational partition function, we can easily obtain the thermodynamic properties for the system as follows:
\begin{enumerate}
\item The vibrational mean energy $U$
\begin{eqnarray}
U(\beta,\zeta)&=&-\frac{\partial}{\partial\beta}1nZ(\beta,\zeta)=\frac{1}{2\beta}\left[1-\frac{\chi}{\mbox{DawsonF}(\chi)}\right]\nonumber\\
 &=&-\frac{2\chi}{\sqrt{\pi}\mbox{Erfi}(\chi)}\left[\frac{e^{\chi^2}}{2\beta}-\frac{\sqrt{\pi}\mbox{Erfi}(\chi)}{4\beta\chi}\right],\ \ \ \ \chi=\chi(\beta, \zeta)=\frac{\zeta}{\tau}\sqrt{\beta},
\label{E55}
\end{eqnarray}
which implies that $U(\beta, \zeta)=-\zeta^2/(3\tau^2)$ when $\beta<<1$.
\item The vibrational specific heat $C$
\begin{eqnarray}
C(\beta,\zeta)&=&\frac{\partial}{\partial T}U(\beta, \zeta)=-k\beta^2\frac{\partial}{\partial\beta}U(\beta, \zeta)\nonumber\\
&=&\frac{1}{2}k\left[1-\frac{\chi\left[2\chi e^{\chi^2}+\sqrt{\pi}\left(1-2\chi^2\right)\mbox{Erfi}(\chi)\right]}{4e^{\chi^2}\mbox{DawsonF}(\chi)^2}\right]
\label{E56}
\end{eqnarray}
which yields $C(\beta, \zeta)=0$ when $\beta<<1$.
\item The vibrational mean free energy $F$. It can be calculated as
\begin{equation}
F(\beta,\zeta)=-kT\ln Z(\beta,\zeta)=-\frac{1}{\beta}\ln\left(\frac{\sqrt{\pi} \tau \mbox{Erfi}\left(\chi\right)}{2 \sqrt{\beta}}\right)
\label{E57}
\end{equation}
\item The vibrational entropy $S$
\begin{eqnarray}
S(\beta,\zeta)&=&k1nZ(\beta,\zeta)+kT\frac{\partial}{\partial T}1nZ_{vib}(\beta,\zeta)\nonumber\\
&=&k1nZ(\beta,\zeta)-k\beta\frac{\partial}{\partial \beta}1nZ(\beta,\zeta)\nonumber\\
&=&\frac{1}{2}k\left[1-\frac{\chi}{\mbox{DawsonF}(\zeta)}+2\log\left(\frac{\tau\mbox{Erfi}(\chi)}{\sqrt{\beta}}\right)+\log\left(\frac{\pi}{4}\right)\right].
\label{E58}
\end{eqnarray}
\end{enumerate}
Here, in this section, we have obtained the thermodynamics properties in terms of two mathemathecal functions namely: the Dawson function and the imaginary error function. In mathematics, the Dawson function or Dawson integral (named for John M. Dawson) can be denoted as \cite{NEW1}
\begin{equation}
F(x)=e^{-x^2}\int_0^x e^{y^2}dy=\frac{\sqrt{\pi}}{2}e^{-x^2}erfi(x).
\end{equation}
Thus the Dawson's integral is implemented in Mathematica as $DawsonF[x]$. On the other hand, the imaginary error function is an entire function defined by
\begin{equation}
erfi(x)=ierf(ix),
\end{equation}
where $erf$ denotes the error function (also called the Gauss error function) is a special function (non-elementary) of sigmoid shape which occurs in probability, statistics and partial differential equations. In mathematics, the error function can be denoted as \cite{NEW1}
\begin{equation}
erf(x)=\frac{2}{\sqrt{\pi}}\int_0^xe^{-t^2}dt.
\end{equation}
The imaginary error function is implemented in Mathematica as Erfi[x].
\section{Numerical Results}
The experimental data of $\mu$ (in amu) and  $\alpha$ (in $\AA^{-1}$) are taken from the recent literature \cite{DAT1} as inputs in expression \ref{E17} to calculate the energy states $E_{n\ell}$(in $eV$) of 12 molecules; namely,  $I_2$, $CO$, $TiH$, $TiC$, $N_2$, $NO$, $CrH$, $NiC$, $O_2$, $LiH$, $VH$, and $ScN$ for different values of the vibrational $n$ and rotational $\ell$ quantum numbers. The experimental data are presented in Table 1. The calculated energy values are listed  Table 2 shows that the PT potential is suitable to describe the diatomic molecules. We selected quantum no $n=0,n=5, n=7$ so as to cover wide energy spectrum in order to see the behavior of energy for large states

In Figure \ref{fig1}, we plot the variation of the vibrational energy levels with the potential parameter $\alpha$ for $ns$ states, where $n=1, 2, 3, 4$. The energy is purely attractive (negative). It is found that energy is increasing in the positive direction with increasing $n$  when $\alpha<0.1$. However, when $\alpha>0.1$, the energy is sharply changing toward the negative direction with the increasing of energy state $n$, i.e., it becomes strongly attractive. For given $\zeta$ and unit of $\tau$, the dependence of the vibrational partition function $Z$ on $\beta$ and $\zeta$ are shown for 4 diatomic molecules; namely, $N_2$, $TiH$, $NiC$ and $I_2$, in Figures \ref{fig2} and \ref{fig3}, respectively. It is found that the $Z$ increases monotonically as $\beta$ and $\zeta$ increase for the first three molecules and linearly increases for the $I_2$ molecule. This means that $I_2$ is not sensitive to the various values $\zeta$ and $\beta$ in Figures \ref{fig2} and \ref{fig3}, respectively. It is shown in Figures \ref{fig4} and \ref{fig5} that the vibrational mean energy $U$ decrease monotonically with the increasing parameters $\beta$ and $\zeta$, respectively. The vibrational specific heat $C$ ($k=1$) first increases with the increasing $\beta$ and $\zeta$ to the maximum value and then decreases with it as shown in Figures \ref{fig6} and \ref{fig7}. In the $I_2$ molecule, $C$  increases monotonically for wide range of $\beta$. The vibrational free energy $F$ increases and then decrases with the increasing parameter $\beta$ in long (short) range for small (large) values of $\zeta$ as shown in Figure \ref{fig8}. On the other hand,  $F$ decreases with the increasing $\zeta$ for various values of $\beta$ and overlapping at a specific value of $\zeta$ as shown in Figure \ref{fig9}. It is shown in Figure \ref{fig10} that the entropy $S(k=1)$ decreases with the parameter $\beta$ for various values of $\zeta$ and overlapping at some specific value $\beta$ in the range. On the other hand, $S$ first increases with the $\zeta$ to the maximum value and then decreases with it as shown in Figure \ref{fig11}. The curves for different $\beta$ are splitting away from each others at higher values of $\zeta$.

\section{Conclusion}
\begin{flushleft}
\begin{figure}[!h]
\includegraphics[height=100mm,width=180mm]{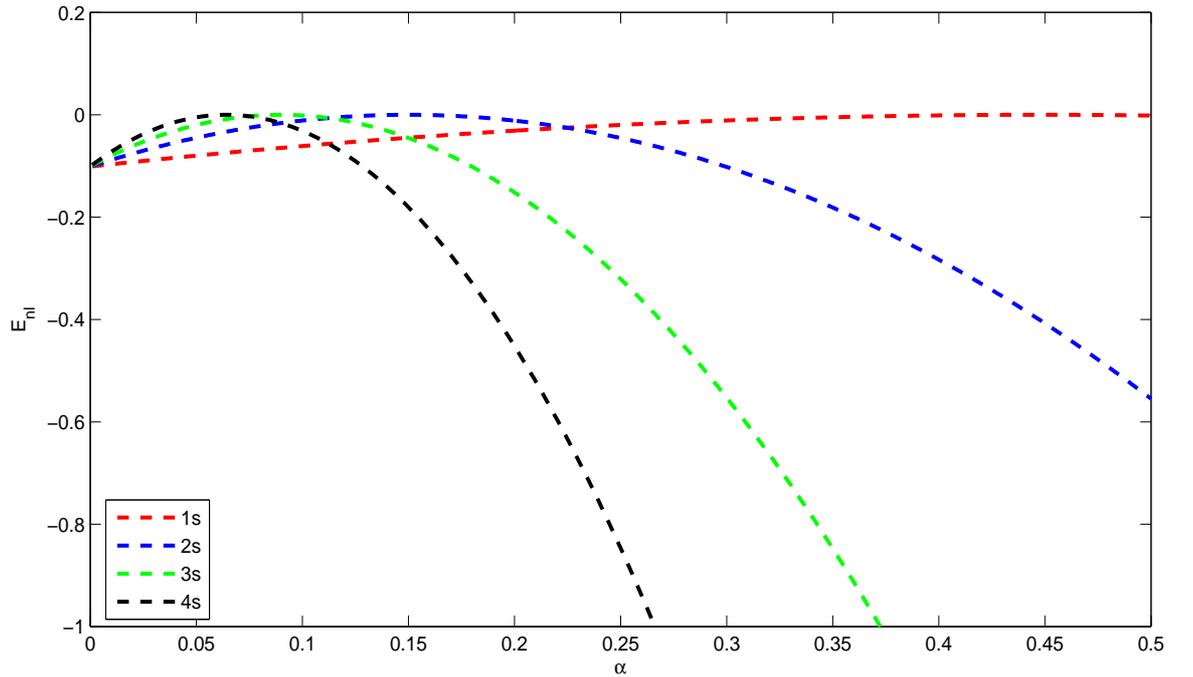}
\caption{{\protect\footnotesize The variation of the energy eigenvalues as a function of potential parameter $\alpha$.}}
\label{fig1}
\end{figure}
\end{flushleft}

\begin{flushleft}
\begin{figure}[!h]
\includegraphics[height=100mm,width=180mm]{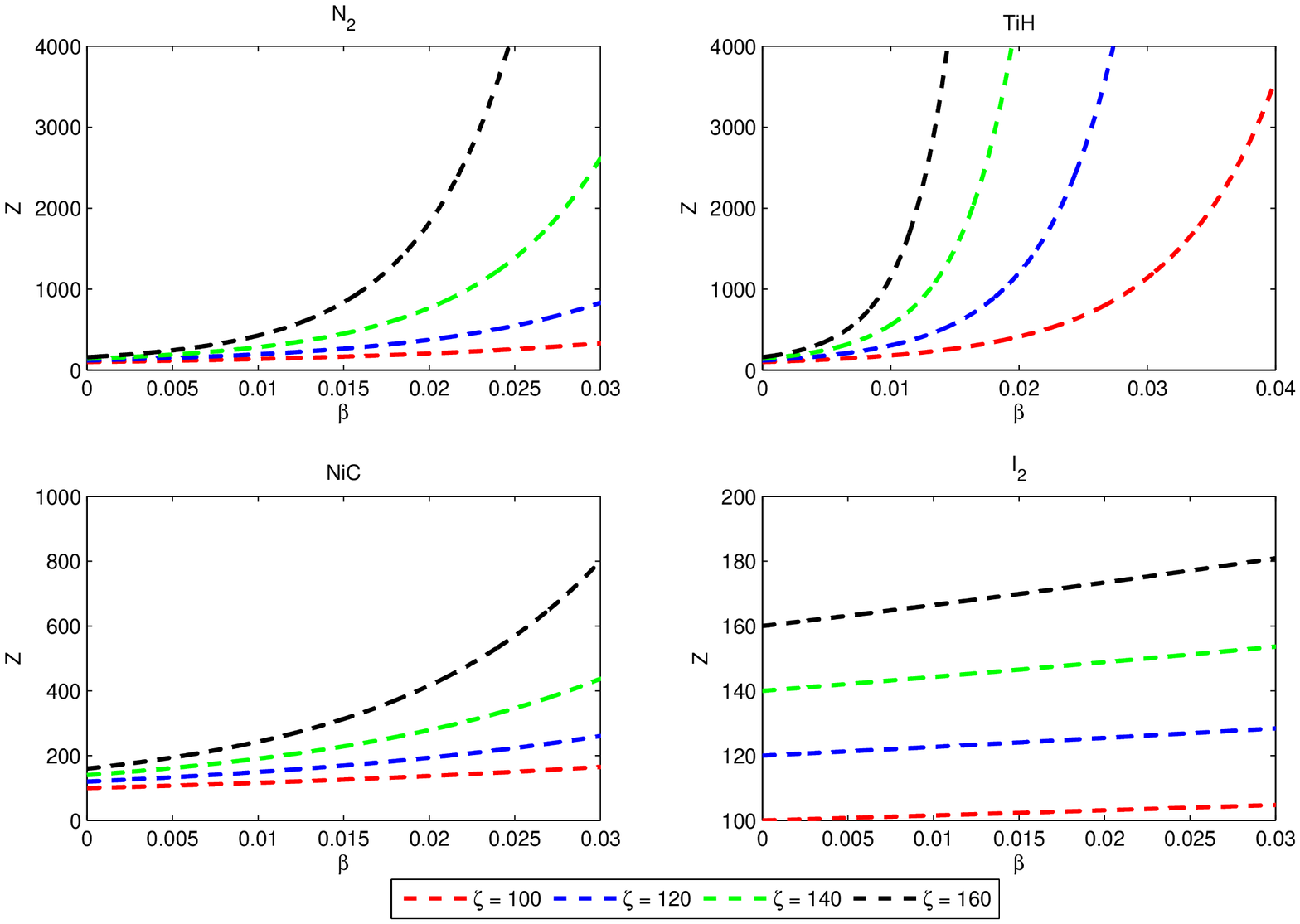}
\caption{{\protect\footnotesize Vibrational partition function $Z$ of $N_2$, $TiH$, $NiC$ and $I_2$ diatomic molecules as a function of $\beta$ for different $\zeta$.}}
\label{fig2}
\end{figure}
\end{flushleft}

\begin{flushleft}
\begin{figure}[!h]
\includegraphics[height=100mm,width=180mm]{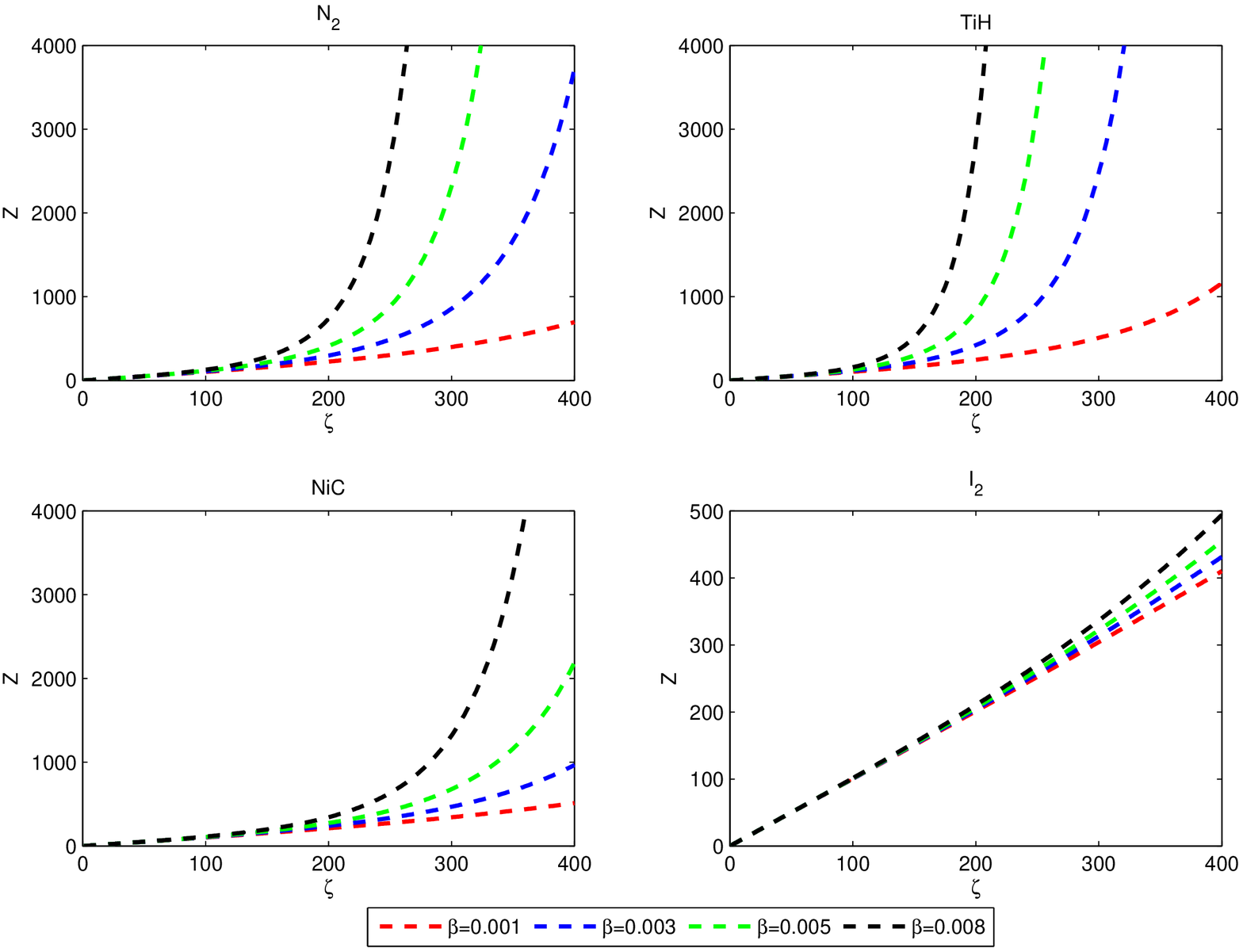}
\caption{{\protect\footnotesize Vibrational partition function $Z$ of $N_2$, $TiH$, $NiC$ and $I_2$ diatomic molecules as a function of  $\zeta$ for different $\beta$.}}
\label{fig3}
\end{figure}
\end{flushleft}

\begin{flushleft}
\begin{figure}[!h]
\includegraphics[height=100mm,width=180mm]{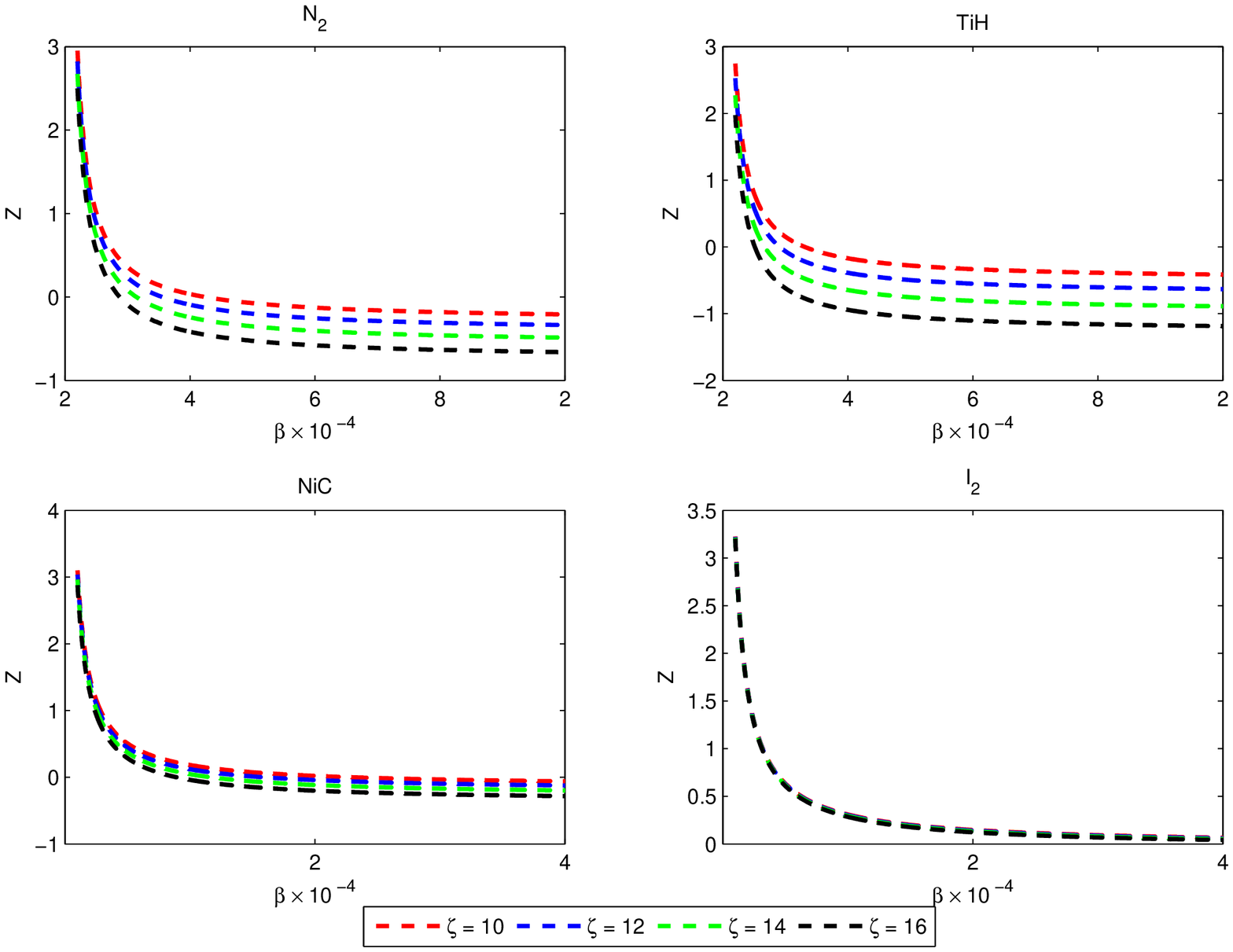}
\caption{{\protect\footnotesize Vibrational mean energy $U$ of $N_2$, $TiH$, $NiC$ and $I_2$ diatomic molecules as a function of $\beta$ for different $\zeta$.}}
\label{fig4}
\end{figure}
\end{flushleft}

\begin{flushleft}
\begin{figure}[!h]
\includegraphics[height=100mm,width=180mm]{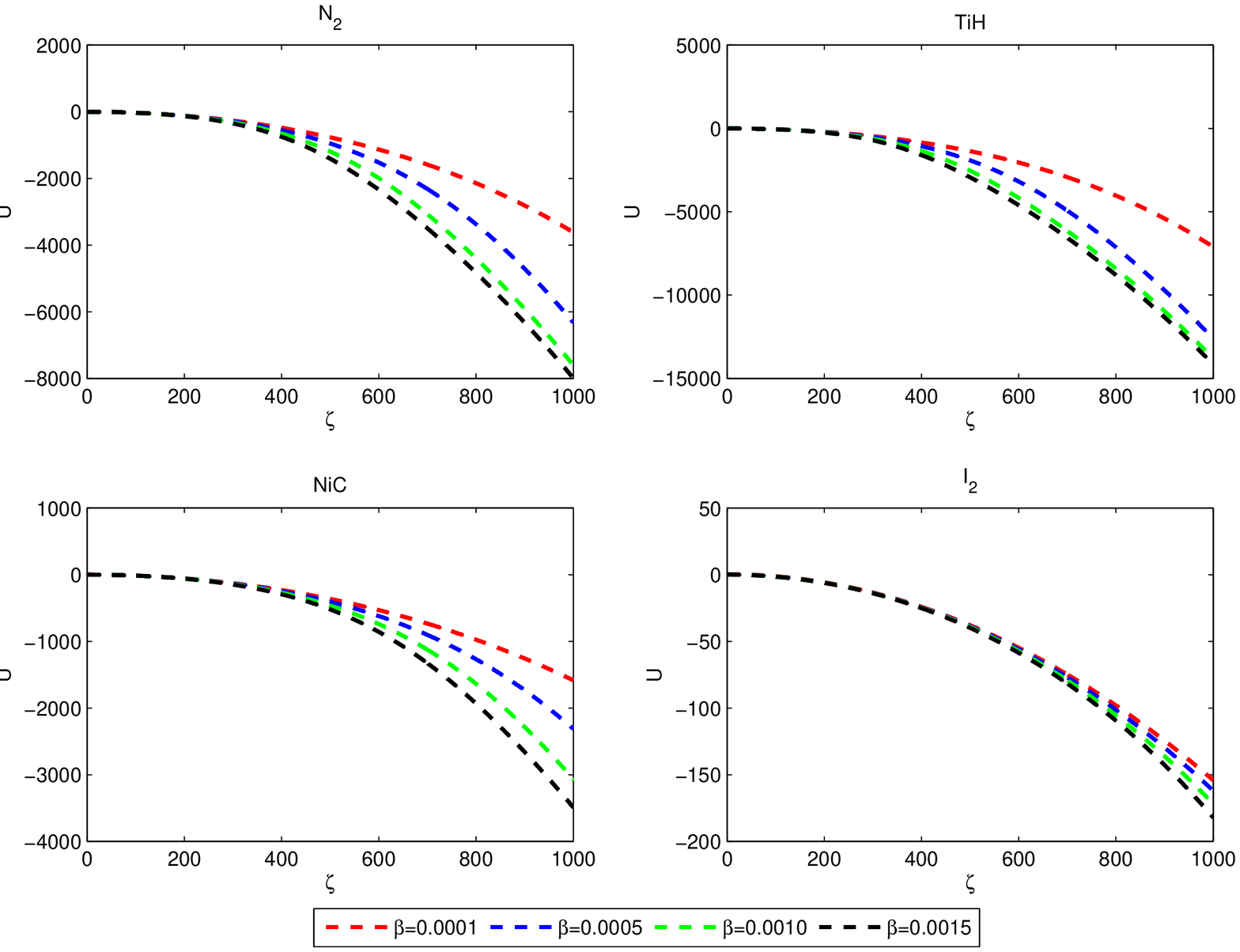}
\caption{{\protect\footnotesize Vibrational mean energy $U$ of $N_2$, $TiH$, $NiC$ and $I_2$ diatomic molecules as a function of  $\zeta$ for different $\beta$.}}
\label{fig5}
\end{figure}
\end{flushleft}

\begin{flushleft}
\begin{figure}[!h]
\includegraphics[height=100mm,width=180mm]{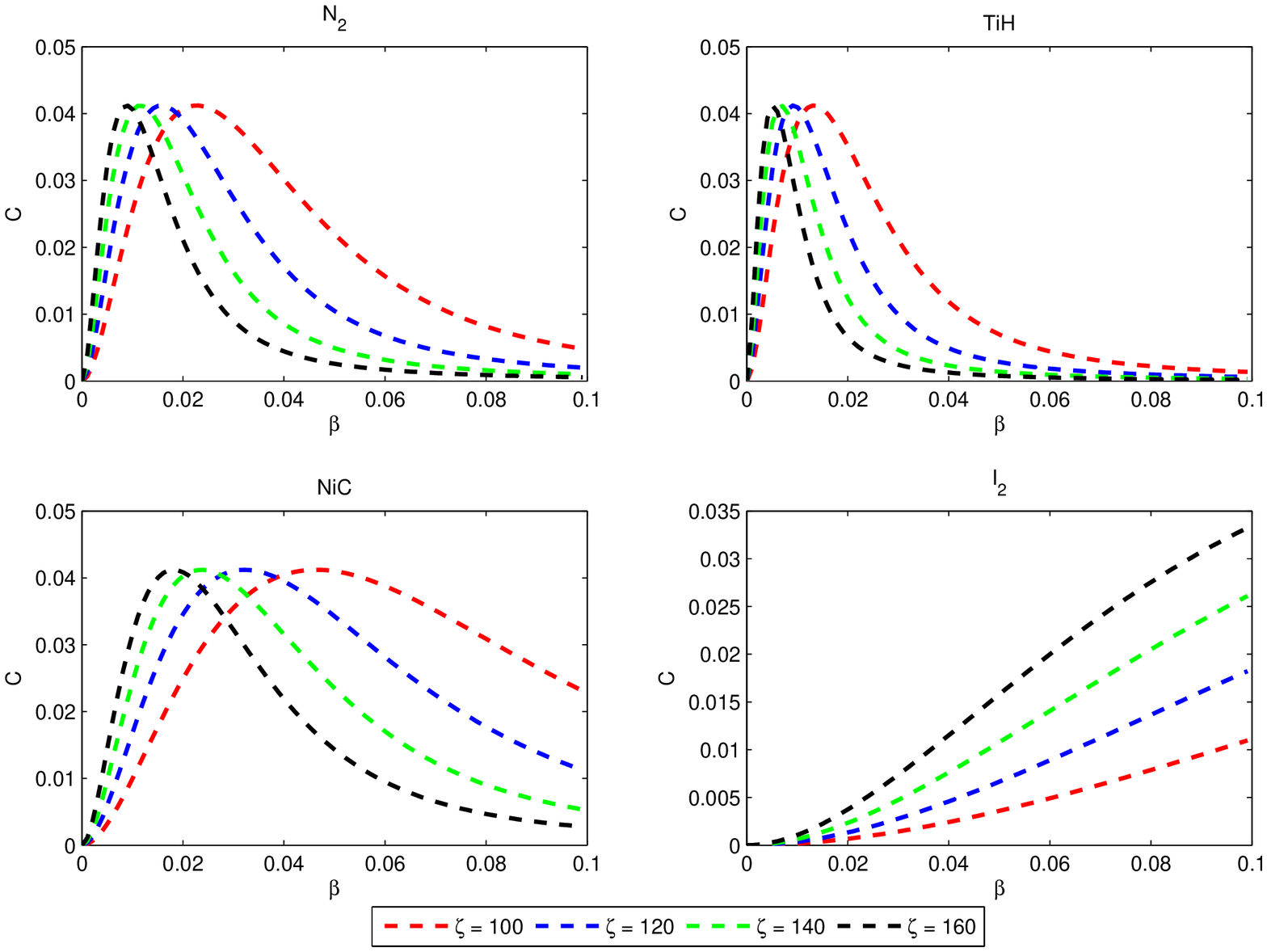}
\caption{{\protect\footnotesize Vibrational specific heat $C$ of $N_2$, $TiH$, $NiC$ and $I_2$ diatomic molecules as a function of $\beta$ for different $\zeta$.}}
\label{fig6}
\end{figure}
\end{flushleft}

\begin{flushleft}
\begin{figure}[!h]
\includegraphics[height=100mm,width=180mm]{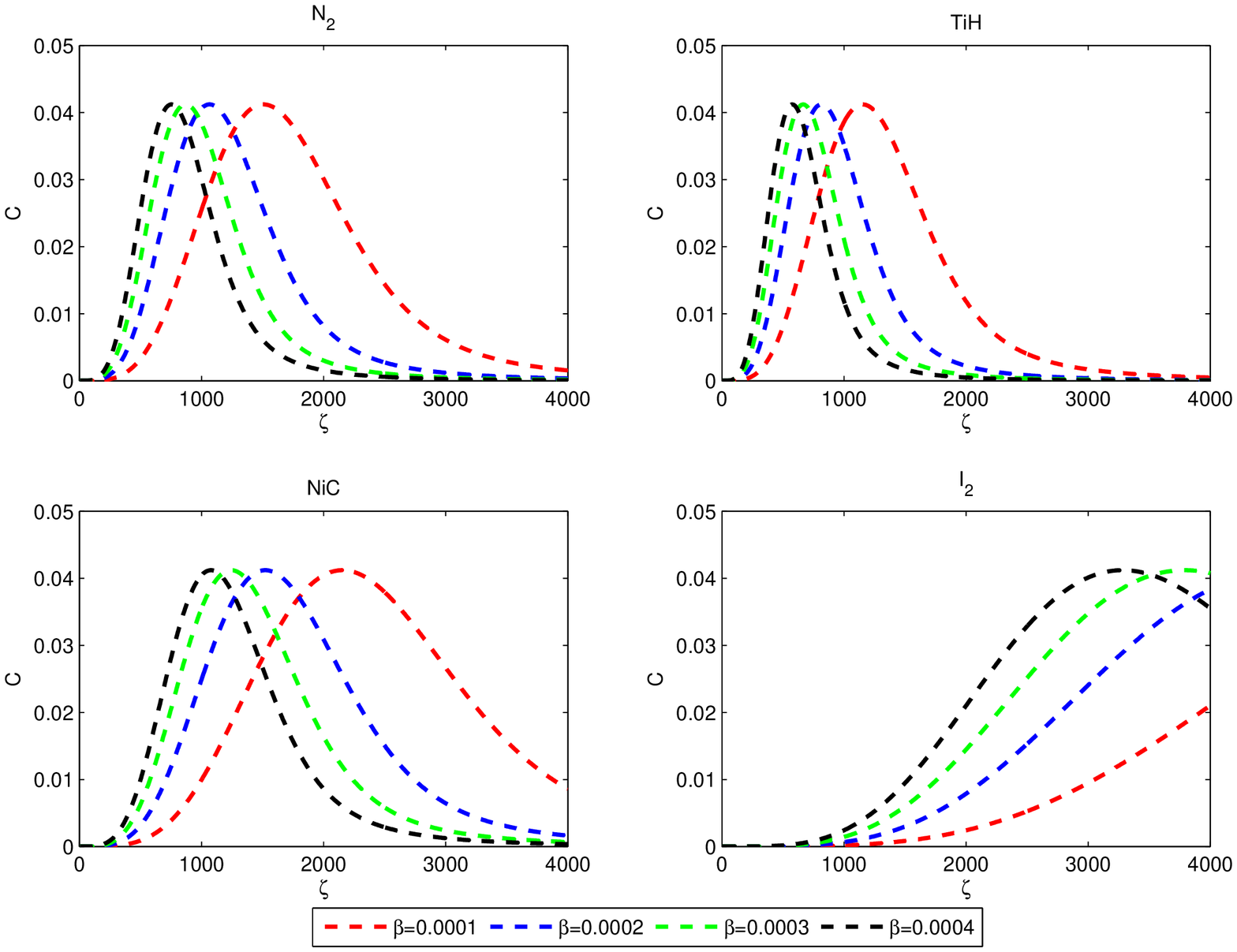}
\caption{{\protect\footnotesize Vibrational specific heat $C$ of $N_2$, $TiH$, $NiC$ and $I_2$ diatomic molecules as a function of  $\zeta$ for different $\beta$.}}
\label{fig7}
\end{figure}
\end{flushleft}

\begin{flushleft}
\begin{figure}[!h]
\includegraphics[height=100mm,width=180mm]{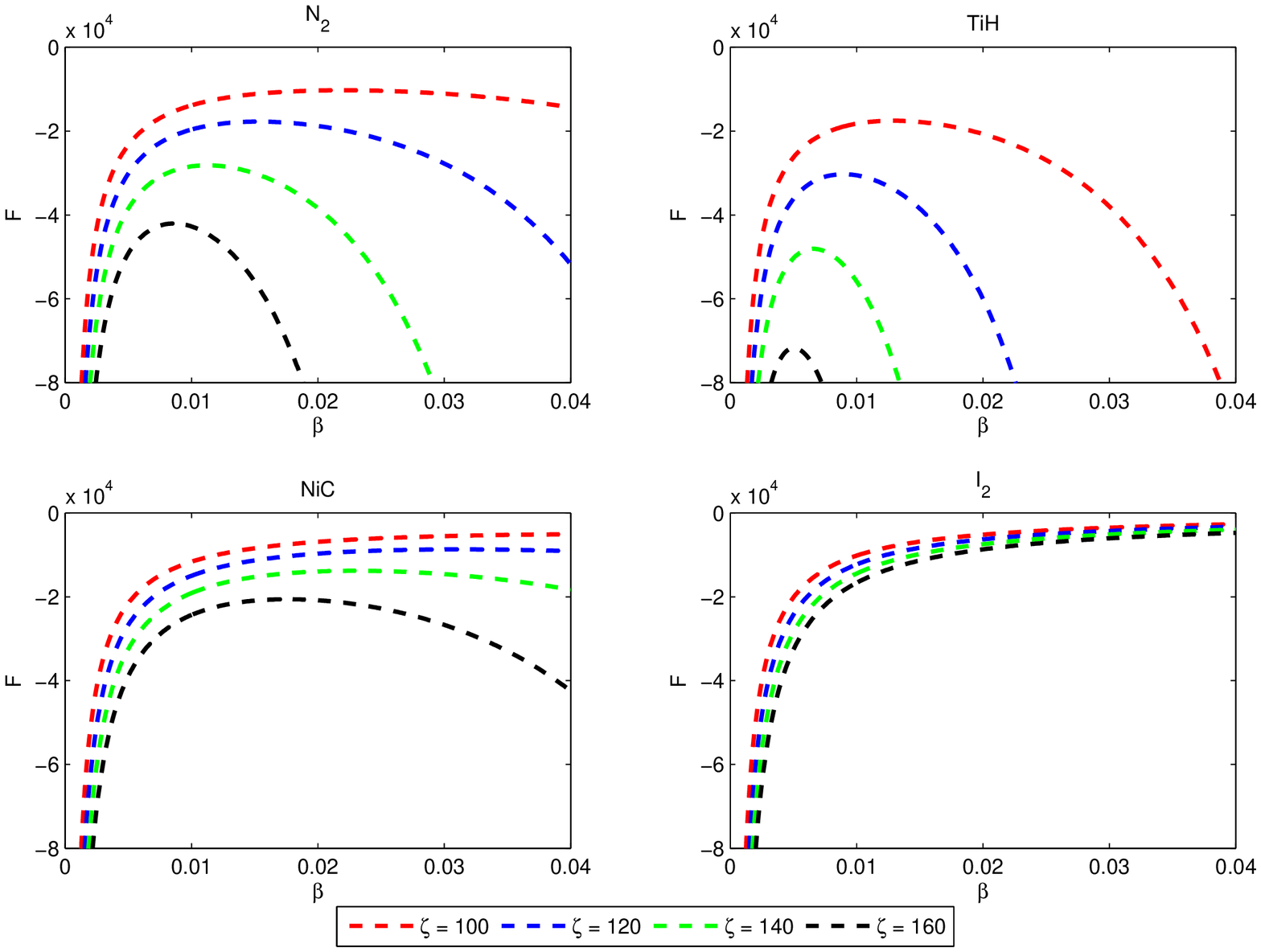}
\caption{{\protect\footnotesize Vibrational free energy $F$ of $N_2$, $TiH$, $NiC$ and $I_2$ diatomic molecules as a function of $\beta$ for different $\zeta$.}}
\label{fig8}
\end{figure}
\end{flushleft}

\begin{flushleft}
\begin{figure}[!h]
\includegraphics[height=100mm,width=180mm]{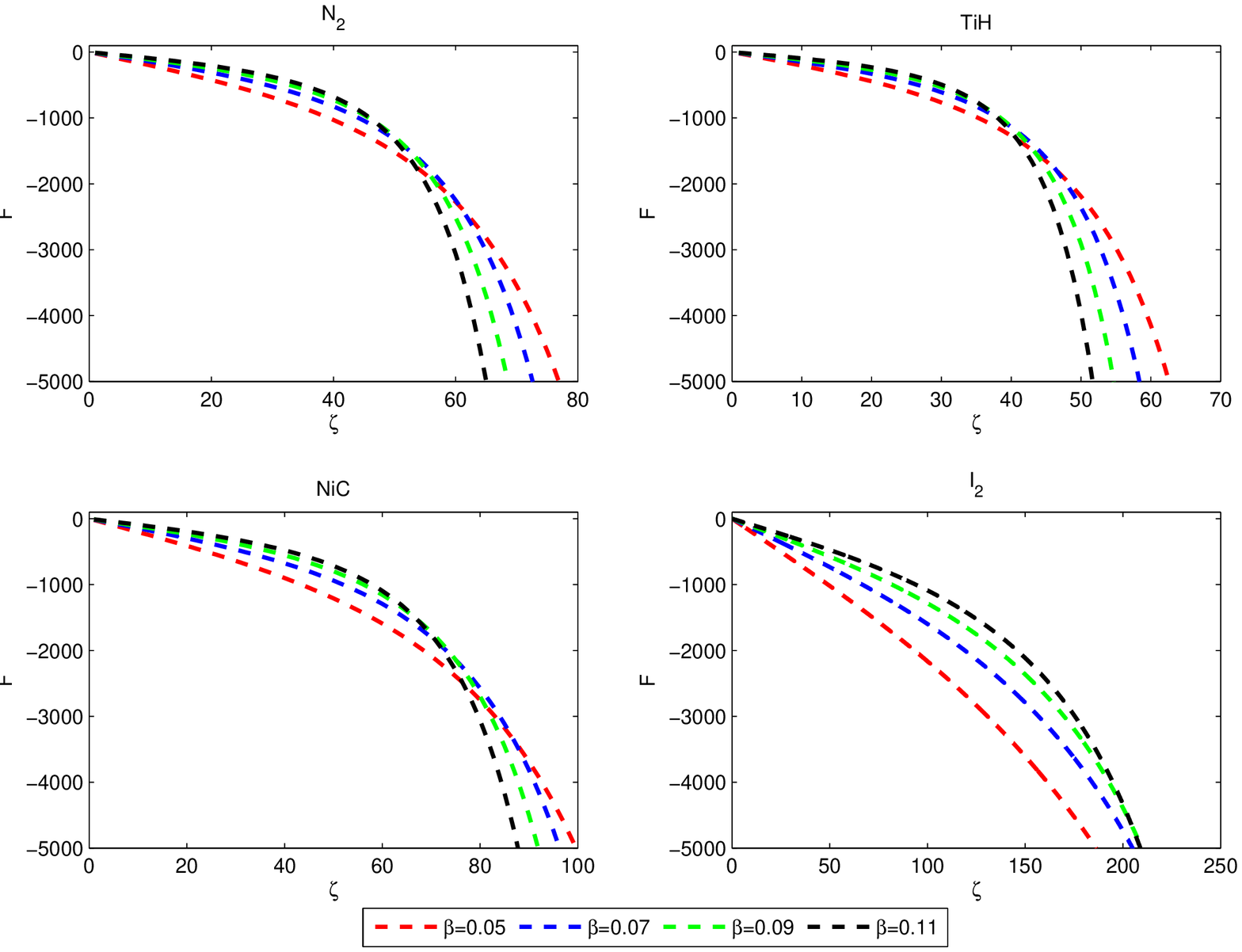}
\caption{{\protect\footnotesize Vibrational free energy $F$ of $N_2$, $TiH$, $NiC$ and $I_2$ diatomic molecules as a function of  $\zeta$ for different $\beta$.}}
\label{fig9}
\end{figure}
\end{flushleft}

\begin{flushleft}
\begin{figure}[!h]
\includegraphics[height=100mm,width=180mm]{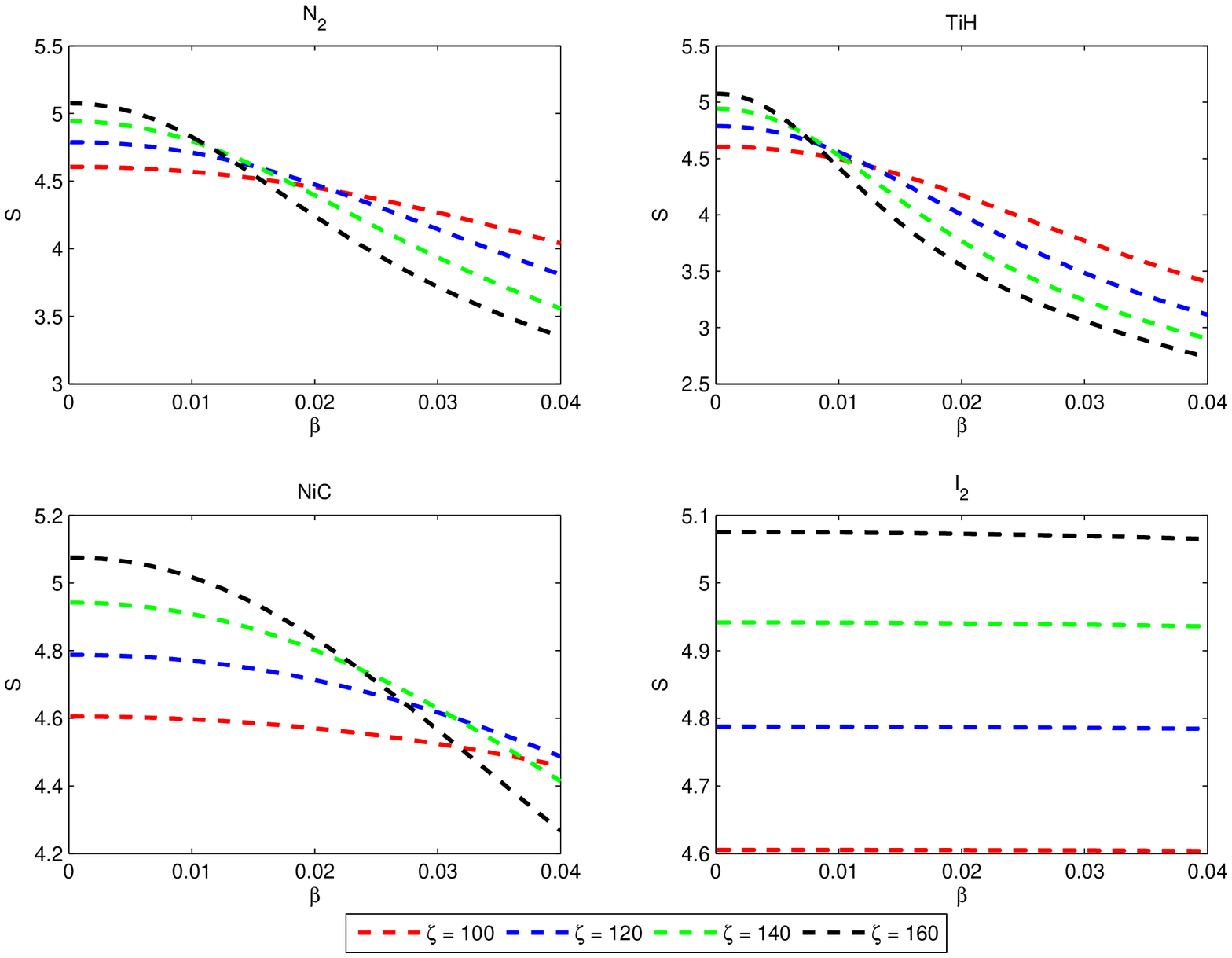}
\caption{{\protect\footnotesize Vibrational mean energy $S$ of $N_2$, $TiH$, $NiC$ and $I_2$ diatomic molecules as a function of $\beta$ for different $\lambda$.}}
\label{fig10}
\end{figure}
\end{flushleft}

\begin{flushleft}
\begin{figure}[!h]
\includegraphics[height=100mm,width=180mm]{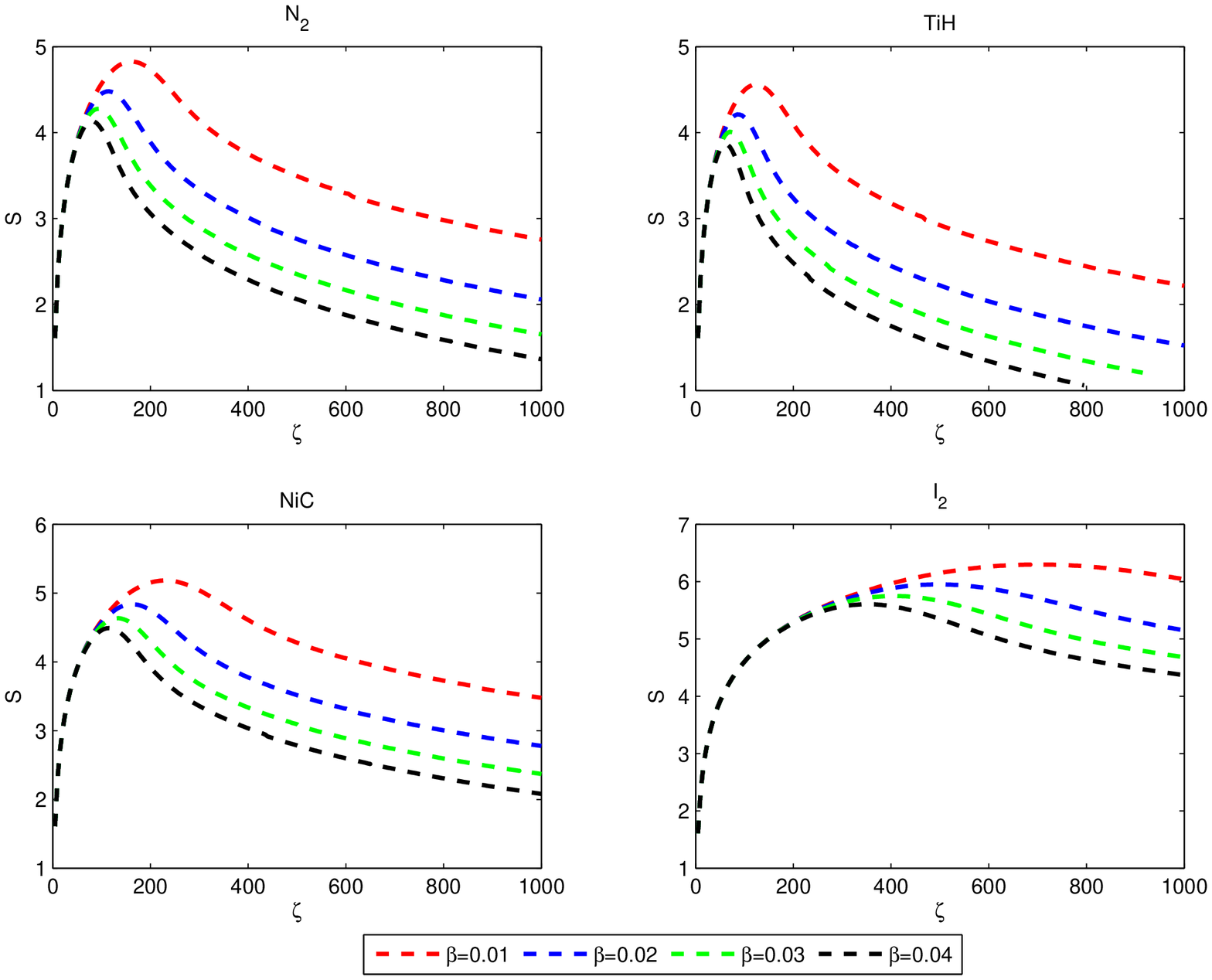}
\caption{{\protect\footnotesize Vibrational free energy $S$ of $N_2$, $TiH$, $NiC$ and $I_2$ diatomic molecules as a function of  $\zeta$ for different $\beta$.}}
\label{fig11}
\end{figure}
\end{flushleft}

In this work, we have solved the Schrodinger equation for the PT potential in the framework of AIM by considering an approximation to deal with the centrifugal term and obtained the energy eigenvalues and the wave functions. The method is also used to obtain approximate energy states and wave functions of the spin-$1/2$ particle in the field of PT potential and Coulomb-like coupling interaction under spin and p-spin symmetries. Some special cases of interest of the present solution are obtained as the $s$ wave case, reflectionless-type potential, symmetric hyperbolic PT potential and the nonrelativistic solution. Further, the nonrelativistic ro-vibrational energy levels of 12 diatomic molecules are obtained using a set parameter values in Tanle 1 for each molecule. Our results are displayed in Table 2.  We plotted the variation of vibrational energy levels with the potential parameter $\alpha$ in Figure \ref{fig1}. On the other hand, we have derived the vibrational partition function $Z$ and then calculated the thermodynamic parameters like the vibrational mean energy $U$, specific heat $C$, free energy $F$ and the entropy $S$. The variations of these thermodynamic parameters with $\beta$ and $\zeta$ are shown in Figures 2-11 for 4 diatomic molecules in presence of PT potential field.  The behaviour of the thermodynamic properties changes from one diatomic molecule to another.

\begin{flushleft}
\textbf{\LARGE Acknowledgements}\\
\end{flushleft}
S. M. Ikhdair thanks the partial support provided by the Scientific and Technical Research Council of Turkey (TUBITAK).

\begin{table}[!h]
{\tiny
\caption{ Model parameters for a few diatomic molecules. \ \ \ \ \ \ \ \ \ \ \ \ \ \ \ \ \ \ \ \ \ \ \ \ \ \ \ \ \ \ \ \ \ \ }\vspace*{10pt}{
\begin{tabular}{ccc}\hline\hline
{}&{}&{}\\[-1.0ex]
Molecules&$\mu$ (in amu)& $\alpha$ (in $\AA^{-1}$)\\[2.5ex]\hline\hline
$I_2$&63.452235020&1.86430\\[1ex]
$CO$ &6.860586000&2.29940\\[1ex]
$TiH$&0.987371000&1.32408\\[1ex]
$TiC$&9.606079000&1.52550\\[1ex]
$N_2$&7.003350000&2.69860\\[1ex]
$NO$ &7.468441000&2.75340\\[1ex]
$CrH$&0.988976000&1.52179\\[1ex]
$NiC$&9.974265000&2.25297\\[1ex]
$O_2$&7.997457504&2.81510\\[1ex]
$LiH$&0.880122100&1.12800\\[1ex]
$VH $&0.988005000&1.44370\\[1ex]
$ScN$&10.68277100&1.50680\\[1ex]\hline\hline
\end{tabular}\label{tab5} }
\vspace*{-1pt}}
\end{table}

\begin{table}[!h]
{\tiny
\caption{Energy spectra $E_{n\ell}$(in $eV$) of 12 diatomic molecules for any arbirary choices of $n$ and $\ell$ quantum numbers with $\hbar c=1973.29eV$$\AA$, $A=-2$ and $B=3$.}\vspace*{10pt}{
\begin{tabular}{cccccccc}\hline\hline
{}&{}&{}&{}&{}&{}&{}&{}\\[-1.0ex]
n&{$\ell$}&$I_2$&$CO$&$TiH$&$TiC$&$N_2$&$NO$\\[2.5ex]\hline\hline
0&	0	 &-2.01518700249	 &-2.05756153769	&-2.08801628475	&-2.03207234893	&-2.06701616738	&-2.06620070795	\\[1ex]
 &	5	 &-1.86617831309	 &-1.52220413799	&-1.30060734002	&-1.72400386632	&-1.45117393674	&-1.45722010494	\\[1ex]
 &	10 &-1.72289229469	 &-1.06736269276	&-0.69870620092	&-1.44124539707	&-0.94397059248 &-0.95429245499	\\[1ex]
5&	0	 &-2.33037239432	 &-3.36982420049	&-4.21935759111	&-2.72413935420	&-3.62461728776	&-3.60232077313	\\[1ex]
 &	5	 &-2.16991836290	 &-2.67343489186	&-3.06093303512	&-2.36545084489	&-2.79149728439	&-2.78123426402	\\[1ex]
 &	10 &-2.01518700249	 &-2.05756153769	&-2.08801628475	&-2.03207234893	&-2.06701616738	&-2.06620070795	\\[1ex]
7&	0	 &-2.46285594258	 &-3.98490713462	&-5.27966285596	&-3.02931337126	&-4.36933328865	&-4.33554810662	\\[1ex]
 &	5	 &-2.29782377436	 &-3.22410506241	&-3.97283205546	&-2.65037685127	&-3.44930217618	&-3.42961923507	\\[1ex]
 &	10 &-2.13851427714	 &-2.54381894467	&-2.85150906060	&-2.29675034463	&-2.63790995008	&-2.62974331656	\\[1ex]
\hline\hline\\[2.5ex]
n&{$\ell$}&$CrH$&$NiC$&$O_2$&$LiH$&$VH$&$ScN$\\[2.5ex]\hline\hline
0&	0	&-2.10140007384	&-2.04665077681	&-2.06539452919	&-2.07925256173	&-2.09612311332	&-2.03002523536	\\[1ex]
 &	5	&-1.20972224650	&-1.6067266228	&-1.46321228291	&-1.36224649240	&-1.24509129115	&-1.74087514222	\\[1ex]
 & 10	&-0.56269024132	&-1.21996969697	&-0.96455612783	&-0.79627951893	&-0.61445809096	&-1.47392957006	\\[1ex]
5&	0	&-4.61869319503	&-3.08600076935	&-3.58033729534	&-3.96638198801	&-4.45938262359	&-2.67493898459	\\[1ex]
 &	5	&-3.23772372335	&-2.53974215899	&-2.77110286666	&-2.94729772693	&-3.16755355746	&-2.34137984949	\\[1ex]
 & 10	&-2.10140007384	&-2.04665077681	&-2.06539452919	&-2.07925256173	&-2.09612311332	&-2.03002523536	\\[1ex]
7&	0	&-5.89961376433	&-3.56128806192	&-4.30226362393	&-4.89039754589	&-5.65153288432	&-2.95777354778	\\[1ex]
 &  5	&-4.32292763492	&-2.97249566902	&-3.41020832231	&-3.75048200812	&-4.18338492061	&-2.60645079589	\\[1ex]
 & 10&-2.99088732768	&-2.4368705043	&-2.62167911187	&-2.76160556622	&-2.93563557888	&-2.27733256498	\\[1ex]\hline\hline
\end{tabular}\label{tab4} }
\vspace*{-1pt}}
\end{table}

\end{document}